\documentclass[fleqn,twoside]{article}
\usepackage{espcrc2}
\usepackage{epsf}

\usepackage{graphicx}
\usepackage[figuresright]{rotating}

\newcommand{\AmS}{{\protect\the\textfont2
  A\kern-.1667em\lower.5ex\hbox{M}\kern-.125emS}}

\title{W Boson Cross Section and Decay Properties at the Tevatron}

\author{Kenneth Bloom\address{University of Michigan, 
for the CDF and D0 Collaborations}}
       
\begin{document}

\begin{abstract}
We present the first measurements of $\sigma(p\bar{p} \to W \to \ell\nu)$ and
$\sigma(p\bar{p} \to Z \to \ell\ell)$ at $\sqrt{s}$ = 1.96 TeV, along
with new measurements of $W$ angular-decay distributions in $p\bar{p}$
collisions at $\sqrt{s}$ = 1.8 TeV.
\vspace{1pc}
\end{abstract}

\maketitle

\section{$W/Z$ production cross sections}

$W$ and $Z$ boson production cross-section measurements in $p\bar{p}$
collisions are a test of the consistency of Standard-Model couplings,
constrain proton parton distribution functions, and provide
information on higher-order QCD corrections.  They also
test the mettle of an experiment, as the measurements require good
understanding of detection efficiencies, backgrounds, and luminosity.
If experimental uncertainties are small, and the cross sections can be
well-estimated from theory, the boson production rates can be
interpreted as a measure of luminosity, and can also be used to
normalize measurements of other production cross sections.  Finally,
$W$ and $Z$ bosons provide a path to the physics of Run II at the
Tevatron, where many signatures of top-quark and Higgs-boson
production can include these bosons.

At the Tevatron, protons and antiprotons collide at a center-of-mass
energy of 1.96 TeV.  A $W$ boson appears in the detector as a high-momentum
lepton and large missing energy due to the undetected neutrino.
As the $z$ component of $p^\nu$ is unmeasured, all quantites are
measured in the transverse plane.  A $Z$ boson appears as two high-momentum
opposite-signed leptons with an invariant mass around 90 GeV.  The
cross section can be expressed as 
\begin{equation}
\sigma \cdot B = \frac{N_{obs} - N_{bg}}{A \epsilon \int {\cal L} dt},
\end{equation}
where $N_{obs}$ is the number of observed boson events, $N_{bg}$ is the
estimated number of background events, $A$ is the kinematic and geometric
acceptance, $\epsilon$ is the total efficiency, and $\int {\cal L} dt$
is the integrated luminosity.

\begin{figure}[htb]
\leavevmode
\epsfxsize=7cm
\epsfbox{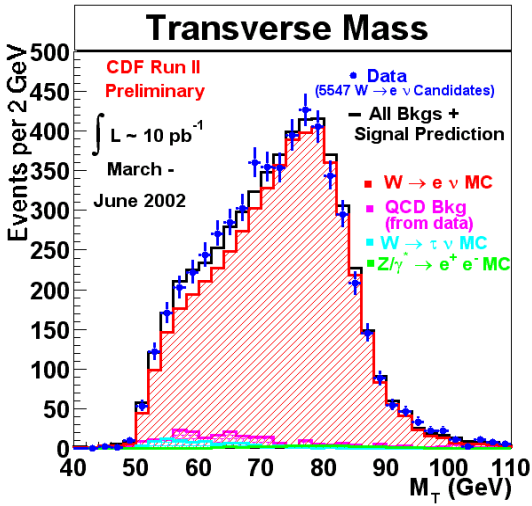}
\caption{Transverse-mass distribution for the CDF electron plus
missing energy sample.}
\label{fig:1}
\end{figure}

The lepton plus missing transverse-energy sample can be accounted for
by $W \to \ell\nu$, $Z \to \ell\ell$, $W \to \tau\nu$, QCD jets (fake
leptons), and (for muons) cosmic rays, as illustrated in
Figure~\ref{fig:1}.  CDF measures the $W$ cross section in the
electron and muon decay channels, and D0 does so in the electron
channel.  (CDF also looks at monojet events with large missing energy,
and sees an enhancement in the number of jets with one and three
charged tracks, evidence for $W \to \tau\nu$ decays.)  The inputs and
results for the cross-section measurements are given in
Table~\ref{tab:1}.  The measurements are systematics-limited, with the
overall uncertainty typically dominated by the uncertainty in the
integrated-luminosity measurement.  The quoted uncertainty on the
luminosity should be considered an upper limit.

\begin{table*}[htb]
\caption{Measurement of the $W$ cross section at CDF and D0.}
\label{tab:1}
\begin{tabular}{l|c|c|c}
& CDF $W \to e\nu$ & CDF $W \to \mu\nu$ 
& D0 $W \to e\nu$\\\hline
$N_{obs}$ & 5547 & 4561 & 9205\\
$N_{bg}$ & 409 $\pm$ 85 & 569 $\pm$ 63 & 5782 $\pm$ 357\\
$A$ (\%) & 23.4 $\pm$ 0.9 & 14.2 $\pm$ 0.4 & 19.6 $\pm$ 0.9\\
$\epsilon$ (\%) & 81.1 $\pm$ 1.8 & 63.2 $\pm$ 3.8 & 86.5 $\pm$ 3.6\\
$\int{\cal L} dt$ (pb$^\mathrm{-1}$) & 10.4 $\pm$ 1.0 &
16.5 $\pm$ 1.6  & 7.5 $\pm$ 0.8\\\hline
$\sigma \cdot B$ (nb) & 
2.60 $\pm$ 0.03$_\mathrm{stat}$ & 2.70 $\pm$ 0.04$_\mathrm{stat}$ 
& 2.67 $\pm$ 0.06$_\mathrm{stat}$\\
& $\pm$ 0.13$_\mathrm{sys}$ $\pm$ 0.26$_\mathrm{lum}$ & 
$\pm$ 0.19$_\mathrm{sys}$ $\pm$ 0.27$_\mathrm{lum}$ &
$\pm$ 0.33$_\mathrm{sys}$ $\pm$ 0.27$_\mathrm{lum}$\\
\end{tabular}
\end{table*}

\begin{figure}[htb]
\epsfxsize=7cm
\epsfbox{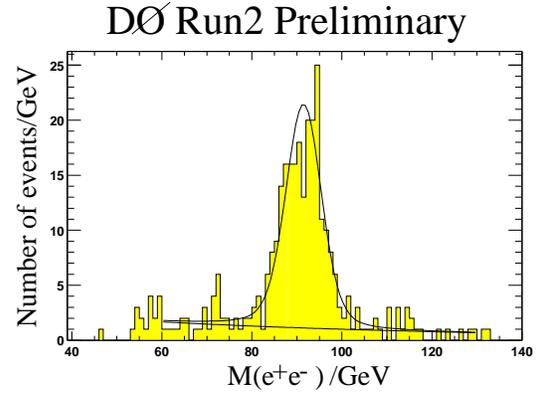}
\caption{Invariant-mass distribution for the D0 dielectron sample.}
\label{fig:2}
\end{figure}

The dilepton samples are quite clean and allow for easy identification
of the $Z$ resonance, as seen in Figure~\ref{fig:2}.  D0 measures
$\sigma \cdot B(Z \to ee)$, with the result shown in
Table~\ref{tab:2}.  From this and the $W$ cross-section measurement,
D0 calculates the ratio of cross sections $R_\ell = \sigma^W/\sigma^Z$,
also given in Table~\ref{tab:2}; CDF measures the same ratio in the muon
channel.  This ratio can be expressed as
\begin{equation}
R_\ell = \frac{\sigma(p\bar{p} \to W)}{\sigma(p\bar{p} \to Z)}
    \frac{\Gamma(W\to \ell\nu)}{\Gamma(Z\to \ell\ell)}
    \frac{\Gamma(Z)}{\Gamma(W)}.
\end{equation}
Taking the total cross-section values and $\Gamma(W\to \ell\nu)$ from
theory, and the $Z$ branching fraction from measurements at the $Z$
pole, $\Gamma(W)$ can be extracted.  The D0 $R$ measurement gives
$\Gamma(W) = 2.26 \pm 0.18_\mathrm{stat} \pm 0.29_\mathrm{sys} \pm
0.04_\mathrm{external}$ GeV, and the CDF measurement of $R$ implies
$\Gamma(W) = 1.67 \pm 0.24^{+0.14}_{-0.13} \pm 0.01$ GeV.  The current
world average is $\Gamma(W) = 2.114 \pm 0.043$ GeV~\cite{PDG}.

\begin{figure}[htb]
\leavevmode
\epsfxsize=7cm
\epsfbox{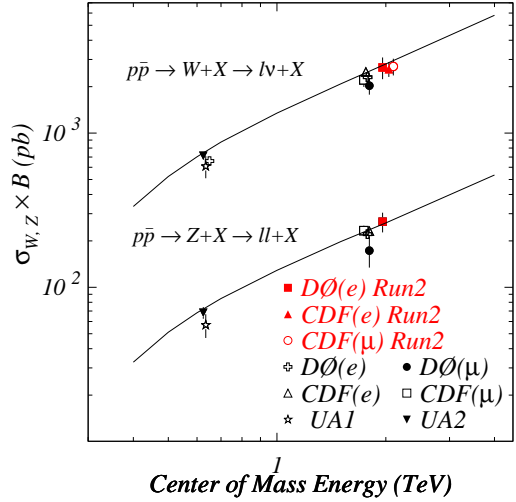}
\caption{$W$ and $Z$ cross-section measurements at different
center-of-mass energies, with theoretical prediction.}
\label{fig:3}
\end{figure}

All of the CDF and D0 cross-section measurements are summarized in
Table~\ref{tab:2}.  For comparison, the best measurements of these
quantities in Run I data are $\sigma^W$ = 2.49 $\pm$ 0.12
nb~\cite{run1W}, $\sigma^Z$ = 249 $\pm$ 11 pb~\cite{run1Z}, and $R$ =
10.90 $\pm$ 0.43~\cite{run1R}.  The Run II measurements have not
reached that level of precision, but they should with additional data.
In addition, the $W$ and $Z$ production cross sections at the Run II
energy of 1.96 TeV are expected to be about 10\% larger than those at
the Run I energy of 1.8 TeV; theory predicts values of $\sigma^W$ =
2.73 (2.50) nb and $\sigma^Z$ = 250 (230) pb at $\sqrt{s}$ = 1.96
(1.8) TeV~\cite{stirling}.  The measured values are consistent with
this predicted rise, as illustrated in Figure~\ref{fig:3}.

\begin{table}[htb]
\caption{Summary of cross-section measurements; uncertainties are
from statistics, systematics, and luminosity normalization.}
\label{tab:2}
\begin{tabular}{l|c|c}
Quantity & Source & Value \\\hline
$\sigma^W_e$ (nb)& CDF & 2.60 $\pm$ 0.03 $\pm$ 0.13 $\pm$ 0.26\\
$\sigma^W_\mu$ (nb)& CDF & 2.70 $\pm$ 0.04 $\pm$ 0.19 $\pm$ 0.27\\
$\sigma^W_e$ (nb)& D0 & 2.67 $\pm$ 0.06 $\pm$ 0.33 $\pm$ 0.27\\\hline
$\sigma^Z_e$ (pb)& D0 & 266 $\pm$ 20 $\pm$ 20 $\pm$ 27\\\hline
$R_e$ & D0 & 10.0 $\pm$ 0.8 $\pm$ 1.3 \\
$R_\mu$ & CDF & 13.7 $\pm 1.9^{+1.1}_{-1.2}$ \\
\end{tabular}
\end{table}

\section{$W$-decay angular distributions}
Understanding QCD effects in $W$ production helps reduce uncertainties
in $M_W$ and other electroweak measurements.  Without QCD, when the
$W$ has no transverse momentum ($p_T^W$), the differential cross
section for the charged lepton in the $W$ decay is $\propto (1 + q_\ell
\cos\theta)^2$, as predicted by the $V-A$ interaction.  But in NLO QCD,
the differential cross section in the Collins-Soper rest frame is 
given as
\begin{eqnarray}
\label{eq}
\frac{d^4\sigma}{dp_T^W dy d(\cos\theta) d\phi} & 
\propto & 1 + \cos^2 \theta + \nonumber\\ 
&& A_0 (1 - 3 \cos^2 \theta)/2 + \nonumber\\
&& A_1 \sin2\theta \cos\phi + \nonumber\\
&& A_2 (\sin^2\theta \cos2\phi)/2 + \nonumber\\
&& A_3 \sin\theta \cos\phi + \nonumber\\
&& A_4 q_\ell \cos\theta + \nonumber\\
&& A_5 \sin^2\theta \cos2\phi + \nonumber\\
&& A_6 \sin2\theta \sin\phi + \nonumber\\
&& A_7 \sin\theta \sin\phi, 
\end{eqnarray}
where the $A_i$ depend on $p_T^W$ and the boson rapidity; the latter
is typically integrated out.  (The $A_1$, $A_5$, $A_6$ and $A_7$ terms
can be safely neglected.)  Thus, angular distributions when $p_T^W
\neq 0$ probe the values of the $A_i$, which can then be compared to
predictions from NLO QCD.  The D0 experiment has made a measurement of
$A_0$ in Run I data\cite{D0A0}, and CDF now has measurements of $A_0$,
$A_2$ and $A_3$ from Run I data.

\begin{figure}[htb]
\leavevmode
\epsfxsize=7cm
\epsfbox{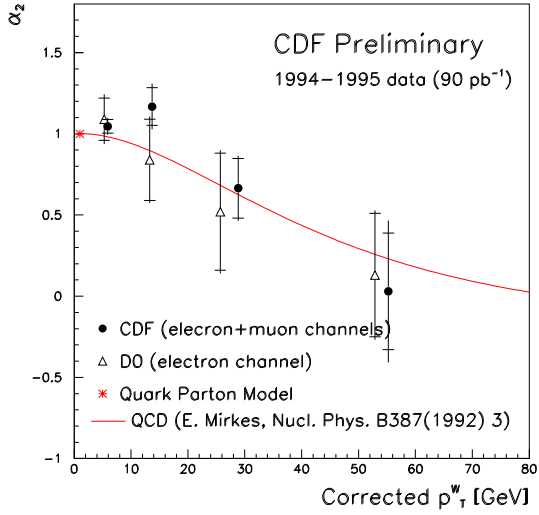}
\caption{CDF measurement of $\alpha_2$ as a function of $p_T^W$, along
with earlier D0 measurement and QCD prediction.}
\label{fig:4}
\end{figure}

The $\theta$ distribution of the decay leptons is obtained by
integrating Equation~\ref{eq} over $\phi$:
\begin{equation}
\frac{d\sigma}{d(\cos\theta)} \propto 1 + q_\ell \alpha_1 \cos\theta + 
\alpha_2 \cos^2\theta,
\end{equation}
with $\alpha_1 = A_4/(2 + A_0)$, $\alpha_2 = (2 - 3A_0)/(2 + A_0)$.
There is no sensitivity to $\alpha_1$ due to the unknown $p^\nu_z$,
but there is sensitivity to $\alpha_2$ through the shape of the $W$
transverse-mass spectrum, which changes with $p_T^W$.  The Run I CDF
$W$ sample is separated into four bins of $p_T^W$, and in each one a
maximum-likelihood fit to the $M_T$ spectrum is performed to extract
$\alpha_2$, accounting for background contamination and detector
acceptance.  The results are shown in Figure~\ref{fig:4}, along with
the earlier D0 results and the prediction from QCD.  There is good
agreement among all three; the decrease in $\alpha_2$ with $p_T^W$
indicates increasing longitudinal polarization, as is expected.  The
CDF measurement is statistics-limited, with systematic uncertainties
dominated by uncertainty in $M_W$, $p_z^W$, and the model of the $W$
recoil.

\begin{figure}[htb]
\leavevmode
\epsfxsize=7cm
\epsfbox{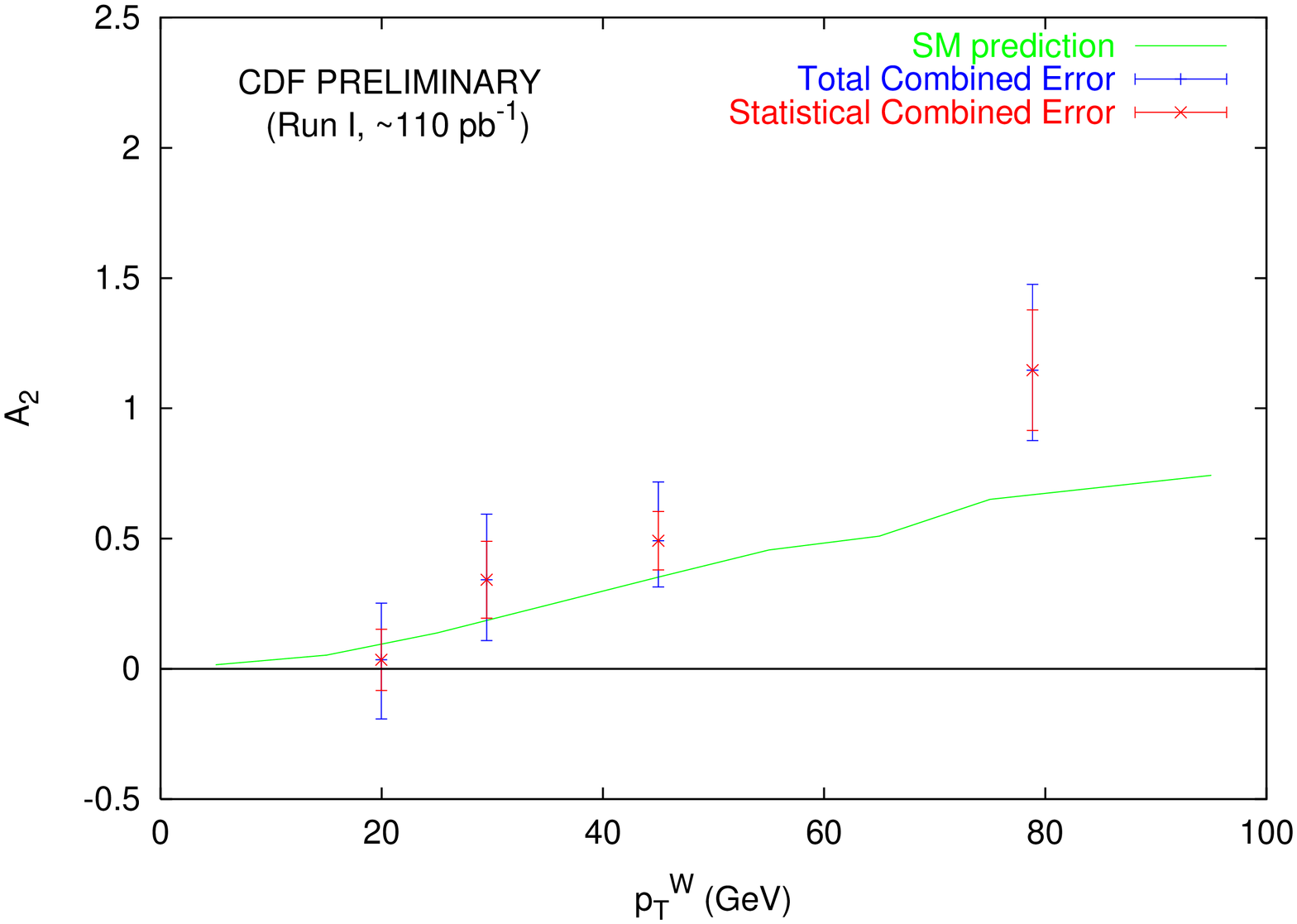}
\newline
\epsfxsize=7cm
\epsfbox{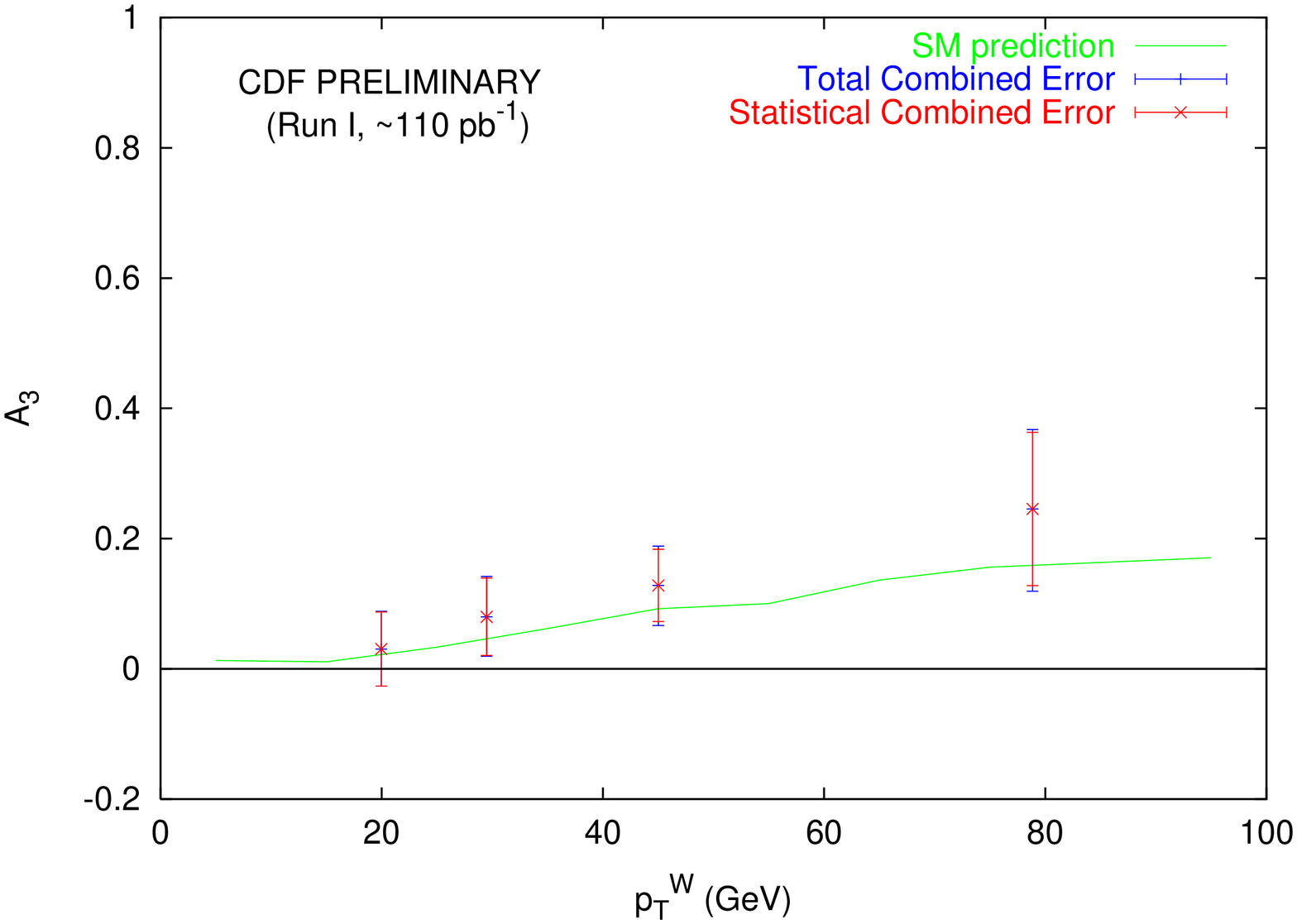}
\caption{CDF measurement of $A_2$ (top) and $A_3$ (bottom) as a
function of $p_T^W$, along with the QCD prediction.}
\label{fig:5}
\end{figure}

Similarly, one can integrate over $\theta$ in Equation~\ref{eq} to
obtain an expression that depends on $\phi$.  Measuring the $\phi$
distribution of the lepton from $W$ decay gives information on the
$A_2$ and $A_3$ coefficients.  Again, events from the Run I CDF $W$
sample are separated into four bins of $p_T^W$; the events are
required to have at least one hadronic jet to ensure that $p_T^W > 0$.
A maximum-likelihood fit to the $\phi$ distribution is performed to
extract $A_2$ and $A_3$.  The results are shown in Figure~\ref{fig:5},
along with the prediction from QCD.  As before, the experimental
results are in agreement with the theory prediction.  The $A_2$
measurement is systematics-limited, with the dominant uncertainties
coming from the knowledge of $A_0$ and $A_4$, and the renormalization
and factorization scale.  By contrast, the $A_3$ measurement is
statistics-limited.

\section{Summary}
Run I at the Tevatron has produced a trove of knowledge about the $W$
and $Z$ bosons, as evidenced by the new CDF measurement of the $W$
angular-decay distributions.  These sophisticated analyses will be
completed soon.  The CDF and D0 collaborations are now turning their
attention to the Run II data, and their $W$ and $Z$-based analyses are
underway.  Both experiments have established techniques to identify
the bosons with high efficiency.  They have a good understanding of
the acceptances and efficiencies of their detector, and this
understanding will improve as more data are collected.  The $W$
cross-section measurements are consistent with the theory predictions
and expectations from earlier measurements, and are already
systematics-limited.  Measurements of the $Z$ cross section and $R$
tend to be statistics-limited, but that too will change as the
datasets grow.  This work provides a solid foundation for the studies
of high-$p_T$ physics that are among the goals of Run II.

\end{document}